\documentclass[%
 reprint,
 amsmath,amssymb,
 aps,prl,superscriptaddress
]{revtex4-1}

\usepackage{graphicx}
\usepackage{dcolumn}
\usepackage{bm}
\usepackage[mathlines]{lineno}

\begin{document}

\preprint{APS/123-QED}

\title{The $^{7}$Be($\boldsymbol{n,\alpha}$)$^{4}$He reaction and the Cosmological Lithium Problem: measurement of the cross section in a wide energy range at n\_TOF (CERN)}

\author{M.~Barbagallo} \affiliation{INFN, Sezione di Bari, Italy}
\author{A.~Musumarra} \affiliation{Dipartimento di Fisica e Astronomia DFA, Universit\`a di Catania, Italy} \affiliation {INFN -  Laboratori Nazionali del Sud, Catania, Italy}
\author{L.~Cosentino} \affiliation{INFN - Laboratori Nazionali del Sud, Catania, Italy}
\author{E.~Maugeri} \affiliation{Paul Scherrer Institut, 5232 Villigen PSI, Switzerland}
\author{S.~Heinitz} \affiliation{Paul Scherrer Institut, 5232 Villigen PSI, Switzerland}
\author{A.~Mengoni}\affiliation{ENEA, Bologna, Italy}
\author{R.~Dressler} \affiliation{Paul Scherrer Institut, 5232 Villigen PSI, Switzerland}
\author{D.~Schumann} \affiliation{Paul Scherrer Institut, 5232 Villigen PSI, Switzerland}
\author{F.~K\"{a}ppeler} \affiliation{Karlsruhe Institute of Technology (KIT), Institut f\"{u}r Kernphysik, Karlsruhe, Germany}
\author{N.~Colonna} \email[]{nicola.colonna@ba.infn.it}\affiliation{INFN, Sezione di Bari, Italy}
\author{P.~Finocchiaro} \affiliation{INFN - Laboratori Nazionali del Sud, Catania, Italy}
\author{M. Ayranov} \affiliation{European Commission, DG-Energy, Luxembourg}
\author{L.~Damone} \affiliation{INFN, Sezione di Bari, Italy}
\author{N.~Kivel} \affiliation{Paul Scherrer Institut, 5232 Villigen PSI, Switzerland}
\author{O.~Aberle} \affiliation{CERN, Geneva, Switzerland}
\author{S.~Altstadt} \affiliation{Johann-Wolfgang-Goethe Universit\"{a}t, Frankfurt, Germany}
\author{J.~Andrzejewski} \affiliation{Uniwersytet \L\'{o}dzki, Lodz, Poland}
\author{L.~Audouin} \affiliation{Centre National de la Recherche Scientifique/IN2P3 - IPN, Orsay, France}
\author{M.~Bacak} \affiliation{Atominstitut der \"{O}sterreichischen Universit\"{a}ten, Technische Universit\"{a}t Wien, Austria}
\author{J. Balibrea-Correa}  \affiliation{Centro de Investigaciones Energeticas Medioambientales y Tecnol\'{o}gicas (CIEMAT), Madrid, Spain}
\author{S.~Barros}\affiliation{C2TN-Instituto Superior Tecn\'{i}co, Universidade de Lisboa, Portugal}
\author{V.~B\'{e}cares}  \affiliation{Centro de Investigaciones Energeticas Medioambientales y Tecnol\'{o}gicas (CIEMAT), Madrid, Spain}
\author{F.~Be\v{c}v\'{a}\v{r}} \affiliation{Charles University, Prague, Czech Republic}
\author{C. Beinrucker}\affiliation{Johann-Wolfgang-Goethe Universit\"{a}t, Frankfurt, Germany}
\author{E.~Berthoumieux} \affiliation{CEA/Saclay - IRFU, Gif-sur-Yvette, France}
\author{J.~Billowes} \affiliation{University of Manchester, Oxford Road, Manchester, UK}
\author{D.~Bosnar} \affiliation{Department of Physics, Faculty of Science, University of Zagreb, Croatia}
\author{M.~Brugger} \affiliation{CERN, Geneva, Switzerland}
\author{M. Caama\~{n}o} \affiliation{Universidade de Santiago de Compostela, Spain}
\author{M.~Calviani} \affiliation{CERN, Geneva, Switzerland}
\author{F.~Calvi\~{n}o} \affiliation{Universitat Politecnica de Catalunya, Barcelona, Spain}
\author{D.~Cano-Ott} \affiliation{Centro de Investigaciones Energeticas Medioambientales y Tecnol\'{o}gicas (CIEMAT), Madrid, Spain}
\author{R.~Cardella} \affiliation{INFN - Laboratori Nazionali del Sud, Catania, Italy} \affiliation{CERN, Geneva, Switzerland}
\author{A. Casanovas} \affiliation{Universitat Politecnica de Catalunya, Barcelona, Spain}
\author{D. M. Castelluccio} \affiliation{ENEA, Bologna, Italy} \
\author{F.~Cerutti} \affiliation{CERN, Geneva, Switzerland}
\author{Y.H.~Chen} \affiliation{Centre National de la Recherche Scientifique/IN2P3 - IPN, Orsay, France}
\author{E.~Chiaveri} \affiliation{CERN, Geneva, Switzerland}
\author{G.~Cort\'{e}s} \affiliation{Universitat Politecnica de Catalunya, Barcelona, Spain}
\author{M.A.~Cort\'{e}s-Giraldo} \affiliation{Universidad de Sevilla, Spain}
\author{M.~Diakaki} \affiliation{National Technical University of Athens (NTUA), Greece}
\author{C.~Domingo-Pardo} \affiliation{Instituto de F{\'{\i}}sica Corpuscular, CSIC-Universidad de Valencia, Spain}
\author{E.~Dupont} \affiliation{CEA/Saclay - IRFU, Gif-sur-Yvette, France}
\author{I.~Duran} \affiliation{Universidade de Santiago de Compostela, Spain}
\author{B. Fernandez-Dominguez} \affiliation{Universidade de Santiago de Compostela, Spain}
\author{A.~Ferrari} \affiliation{CERN, Geneva, Switzerland}
\author{P. Ferreira}\affiliation{C2TN-Instituto Superior Tecn\'{i}co, Universidade de Lisboa, Portugal}
\author{W. Furman} \affiliation{Joint Institute of Nuclear Research, Dubna, Russia}
\author{S.~Ganesan}\affiliation{Bhabha Atomic Research Centre (BARC), Mumbai, India}
\author{A.~Garc{\'{\i}}a-Rios} \affiliation{Centro de Investigaciones Energeticas Medioambientales y Tecnol\'{o}gicas (CIEMAT), Madrid, Spain}
\author{A. Gawlik} \affiliation{Uniwersytet \L\'{o}dzki, Lodz, Poland}
\author{T. Glodariu} \affiliation{Horia Hulubei National Institute of Physics and Nuclear Engineering - IFIN HH, Bucharest - Magurele, Romania}
\author{K. G\"{o}bel}\affiliation{Johann-Wolfgang-Goethe Universit\"{a}t, Frankfurt, Germany}
\author{I.F.~Gon\c{c}alves}\affiliation{C2TN-Instituto Superior Tecn\'{i}co, Universidade de Lisboa, Portugal}
\author{E.~Gonz\'{a}lez-Romero} \affiliation{Centro de Investigaciones Energeticas Medioambientales y Tecnol\'{o}gicas (CIEMAT), Madrid, Spain}
\author{E.~Griesmayer} \affiliation{Atominstitut der \"{O}sterreichischen Universit\"{a}ten, Technische Universit\"{a}t Wien, Austria}
\author{C.~Guerrero} \affiliation{Universidad de Sevilla, Spain}
\author{F.~Gunsing} \affiliation{CEA/Saclay - IRFU, Gif-sur-Yvette, France}
\author{H. Harada} \affiliation{Japan Atomic Energy Agency (JAEA), Tokai-mura, Japan}
\author{T. Heftrich}\affiliation{Johann-Wolfgang-Goethe Universit\"{a}t, Frankfurt, Germany}
\author{J. Heyse} \affiliation{European Commission JRC, Institute for Reference Materials and Measurements, Retieseweg 111, B-2440 Geel, Belgium}
\author{D.G.~Jenkins} \affiliation{University of York, Heslington, York, UK}
\author{E.~Jericha} \affiliation{Atominstitut der \"{O}sterreichischen Universit\"{a}ten, Technische Universit\"{a}t Wien, Austria}
\author{T. Katabuchi} \affiliation{Tokyo Institute of Technology, Japan}
\author{P. Kavrigin} \affiliation{Atominstitut der \"{O}sterreichischen Universit\"{a}ten, Technische Universit\"{a}t Wien, Austria}
\author{A. Kimura} \affiliation{Japan Atomic Energy Agency (JAEA), Tokai-mura, Japan}
\author{M.~Kokkoris} \affiliation{National Technical University of Athens (NTUA), Greece}
\author{M.~Krti\v{c}ka} \affiliation{Charles University, Prague, Czech Republic}
\author{E. Leal-Cidoncha} \affiliation{Universidade de Santiago de Compostela, Spain}
\author{J. Lerendegui} \affiliation{Universidad de Sevilla, Spain}
\author{C.~Lederer} \affiliation{School of Physics and Astronomy, University of Edinburgh, UK}
\author{H.~Leeb} \affiliation{Atominstitut der \"{O}sterreichischen Universit\"{a}ten, Technis
che Universit\"{a}t Wien, Austria}
\author{S.~Lo Meo} \affiliation{ENEA, Bologna, Italy} \affiliation{INFN, Sezione di Bologna, Italy}
\author{S.J. Lonsdale} \affiliation{School of Physics and Astronomy, University of Edinburgh, UK}
\author{R.~Losito} \affiliation{CERN, Geneva, Switzerland}
\author{D.~Macina} \affiliation{CERN, Geneva, Switzerland}
\author{J.~Marganiec} \affiliation{Uniwersytet \L\'{o}dzki, Lodz, Poland}
\author{T.~Mart\'{\i}nez} \affiliation{Centro de Investigaciones Energeticas Medioambientales y Tecnol\'{o}gicas (CIEMAT), Madrid, Spain}
\author{C.~Massimi} \affiliation{Dipartimento di Fisica e Astronomia, Universit\`a di Bologna} \affiliation{INFN, Sezione di Bologna, Italy}
\author{P.~Mastinu} \affiliation{INFN - Laboratori Nazionali di Legnaro, Italy}
\author{M.~Mastromarco} \affiliation{INFN, Sezione di Bari, Italy}
\author{A.~Mazzone} \affiliation{CNR - IC, Bari, Italy} \affiliation{INFN, Sezione di Bari, Italy}
\author{E.~Mendoza} \affiliation{Centro de Investigaciones Energeticas Medioambientales y Tecnol\'{o}gicas (CIEMAT), Madrid, Spain}
\author{P.M.~Milazzo} \affiliation{INFN, Sezione di Trieste, Italy}
\author{F.~Mingrone} \affiliation{Dipartimento di Fisica e Astronomia, Universit\`a di Bologna} \affiliation{INFN, Sezione di Bologna, Italy}
\author{M.~Mirea} \affiliation{Horia Hulubei National Institute of Physics and Nuclear Engineering - IFIN HH, Bucharest - Magurele, Romania}
\author{S.~Montesano} \affiliation{CERN, Geneva, Switzerland}
\author{R.~Nolte} \affiliation{Physikalisch Technische Bundesanstalt (PTB), Braunschweig, Germany}
\author{A.~Oprea} \affiliation{Horia Hulubei National Institute of Physics and Nuclear Engineering - IFIN HH, Bucharest - Magurele, Romania}
\author{A.~Pappalardo} \affiliation{INFN - Laboratori Nazionali del Sud, Catania, Italy}
\author{N.~Patronis} \affiliation{University of Ioannina, Greece}
\author{A.~Pavlik}\affiliation{University of Vienna, Faculty of Physics, Austria}
\author {J.~Perkowski} \affiliation{Uniwersytet \L\'{o}dzki, Lodz, Poland}
\author{M.~Piscopo} \affiliation{INFN - Laboratori Nazionali del Sud, Catania, Italy}
\author{A.~Plompen}\affiliation{European Commission JRC, Institute for Reference Materials and Measurements, Retieseweg 111, B-2440 Geel, Belgium}
\author {I.~Porras} \affiliation{Universidad de Granada, Spain}
\author {J.~Praena} \affiliation{Universidad de Sevilla, Spain} \affiliation{Universidad de Granada, Spain}
\author{J.~Quesada} \affiliation{Universidad de Sevilla, Spain}
\author{K. Rajeev}\affiliation{Bhabha Atomic Research Centre (BARC), Mumbai, India}
\author{T.~Rauscher} \affiliation{Centre for Astrophysics Research, School of Physics, Astronomy and Mathematics, University of Hertfordshire, Hatfield, UK} \affiliation{Department of Physics, University of Basel, Basel, Switzerland}
\author{R.~Reifarth}\affiliation{Johann-Wolfgang-Goethe Universit\"{a}t, Frankfurt, Germany}
\author{A.~Riego-Perez} \affiliation{Universitat Politecnica de Catalunya, Barcelona, Spain}
\author{P. Rout}\affiliation{Bhabha Atomic Research Centre (BARC), Mumbai, India}
\author{C.~Rubbia}  \affiliation{CERN, Geneva, Switzerland}
\author{J. Ryan} \affiliation{University of Manchester, Oxford Road, Manchester, UK}
\author{M. Sabate-Gilarte} \affiliation{CERN, Geneva, Switzerland}
\author{A.~Saxena} \affiliation{Bhabha Atomic Research Centre (BARC), Mumbai, India}
\author{P.~Schillebeeckx} \affiliation{European Commission JRC, Institute for Reference Materials and Measurements, Retieseweg 111, B-2440 Geel, Belgium}
\author{S.~Schmidt}\affiliation{Johann-Wolfgang-Goethe Universit\"{a}t, Frankfurt, Germany}
\author{P. Sedyshev} \affiliation{Joint Institute of Nuclear Research, Dubna, Russia}
\author{A. G. Smith} \affiliation{University of Manchester, Oxford Road, Manchester, UK}
\author{A. Stamatopoulos} \affiliation{National Technical University of Athens (NTUA), Greece}
\author{G.~Tagliente} \affiliation{INFN, Sezione di Bari, Italy}
\author{J.L.~Tain} \affiliation{Instituto de F{\'{\i}}sica Corpuscular, CSIC-Universidad de Valencia, Spain}
\author{A. Tarife\~{n}o-Saldivia} \affiliation{Instituto de F{\'{\i}}sica Corpuscular, CSIC-Universidad de Valencia, Spain}
\author{L.~Tassan-Got} \affiliation{Centre National de la Recherche Scientifique/IN2P3 - IPN, Orsay, France}
\author{A.~Tsinganis}  \affiliation{CERN, Geneva, Switzerland}
\author{S.~Valenta}\affiliation{Charles University, Prague, Czech Republic}
\author{G.~Vannini} \affiliation{Dipartimento di Fisica e Astronomia, Universit\`a di Bologna} \affiliation{INFN, Sezione di Bologna, Italy}
\author{V.~Variale} \affiliation{INFN, Sezione di Bari, Italy}
\author{P.~Vaz} \affiliation{C2TN-Instituto Superior Tecn\'{i}co, Universidade de Lisboa, Portugal}
\author{A.~Ventura} \affiliation{INFN, Sezione di Bologna, Italy}
\author{V.~Vlachoudis} \affiliation{CERN, Geneva, Switzerland}
\author{R.~Vlastou} \affiliation{National Technical University of Athens (NTUA), Greece}
\author{J. Vollaire} \affiliation{CERN, Geneva, Switzerland}
\author{A.~Wallner} \affiliation{Research School of Physics and Engineering, Australian National University, ACT 0200, Australia} \affiliation{University of Vienna, Faculty of Physics, Austria}
\author{S.~Warren} \affiliation{University of Manchester, Oxford Road, Manchester, UK}
\author{M.~Weigand} \affiliation{Johann-Wolfgang-Goethe Universit\"{a}t, Frankfurt, Germany}
\author{C.~Wei{\ss}} \affiliation{CERN, Geneva, Switzerland}
\author{C. Wolf} \affiliation{Johann-Wolfgang-Goethe Universit\"{a}t, Frankfurt, Germany}
\author{P.J.~Woods} \affiliation{School of Physics and Astronomy, University of Edinburgh, UK}
\author{T.~Wright} \affiliation{University of Manchester, Oxford Road, Manchester, UK}
\author{P.~\v{Z}ugec} \affiliation{Department of Physics, Faculty of Science, University of Zagreb, Croatia}

\collaboration{The n\_TOF Collaboration (www.cern.ch/ntof)}  \noaffiliation

\date{\today}

\begin{abstract}
The energy-dependent cross section of the $^{7}$Be($n,\alpha$)$^{4}$He reaction, of interest for the so-called Cosmological Lithium Problem in Big Bang Nucleosynthesis, has been measured for the first time from 10 meV to 10 keV neutron energy. The challenges posed by the short half-life of $^{7}$Be and by the low reaction cross section have been overcome at n\_TOF thanks to an unprecedented combination of the extremely high luminosity and good resolution of the neutron beam in the new experimental area (EAR2) of the n\_TOF facility at CERN, the availability of a sufficient amount of chemically pure $^{7}$Be, and a specifically designed experimental setup. Coincidences between the two $\alpha$-particles have been recorded in two Si-$^{7}$Be-Si arrays placed directly in the neutron beam. The present results are consistent, at thermal neutron energy, with the only previous measurement performed in the 60's at a nuclear reactor. The energy dependence here reported clearly indicates the inadequacy of the cross section estimates currently used in BBN calculations. Although new measurements at higher neutron energy may still be needed, the n\_TOF results hint to a minor role of this reaction in BBN, leaving the long-standing Cosmological Lithium problem unsolved.

\begin{description}
\item[PACS numbers]
\pacs{}26.35.+c, 28.20.-v, 27.20.+n


\end{description}

\end{abstract}

\maketitle


One of the most important unresolved problems in Nuclear Astrophysics is the so-called Cosmological Lithium problem \cite{CLIP1}, i.e. the large discrepancy between the abundance of primordial $^{7}$Li predicted by the standard theory of Big Bang Nucleosynthesis (BBN) and the value deduced from the observation of galactic halo dwarf stars. In contrast to other isotopes, whose primordial abundances are succesfully reproduced by BBN calculations, $^{7}$Li is overestimated by more than a factor of 3, relative to the value inferred from the so-called Spite plateau halo stars.

In the standard BBN 95\% of the primordial $^{7}$Li is produced by the electron capture decay of $^{7}$Be (t$_{1/2}$=53.2 d), relatively late after the Big Bang, when the Universe has cooled down for electrons and nuclei to combine into atoms. Therefore, the abundance of $^{7}$Li is essentially determined by the production and destruction of $^{7}$Be. Several mechanisms have been put forward to explain the difference between calculations and observations \cite{CLIP1a, CLIP1b, CLIP1c, CLIP1d,izzo}: new physics beyond the Standard Model, errors in the inferred primordial $^{7}$Li abundance from the Spite plateau stars and, finally, systematic uncertainties in the Nuclear Physics used in the BBN calculations, in particular for reactions leading to the destruction of $^{7}$Be. The possibility that charged particle induced reactions could be responsible for the destruction of $^{7}$Be during BBN has been ruled out by previous experiments \cite{CLIP2,CLIP3,CLIP4,broggini}, but neutron-induced reactions on $^{7}$Be need still to be considered. 

In this context, a significant impact of the $^{7}$Be($n, p$) reaction has been excluded by a measurement at the LANSCE neutron facility, Los Alamos \cite{koehler}, but so far the contribution of the $^{7}$Be($n, \alpha$)$^{4}$He channel to the destruction of  $^{7}$Be has been considered negligible in BBN calculations due to the much lower cross-section estimated for this reaction. However, this assumption has never been verified experimentally. In literature, only one measurement of the $^{7}$Be($n, \alpha$)$^{4}$He cross section at thermal neutron energy is reported, performed at the ISPRA reactor in 1963 \cite{bassi}. This result was used for extrapolation of the cross section to BBN energies by Wagoner in 1967 \cite{wagoner}. Other theoretical calculations in the keV neutron energy region yield completely different results, with discrepancies of up to two orders of magnitude \cite{hou}. A cross section for the ($n, \alpha$) reaction two orders of magnitude higher than currently used in BBN calculations, in the pertinent neutron energy region, could solve the Cosmological Lithium problem \cite{broggini}. 

\begin{figure}[t]
  \includegraphics[width=8.5cm]{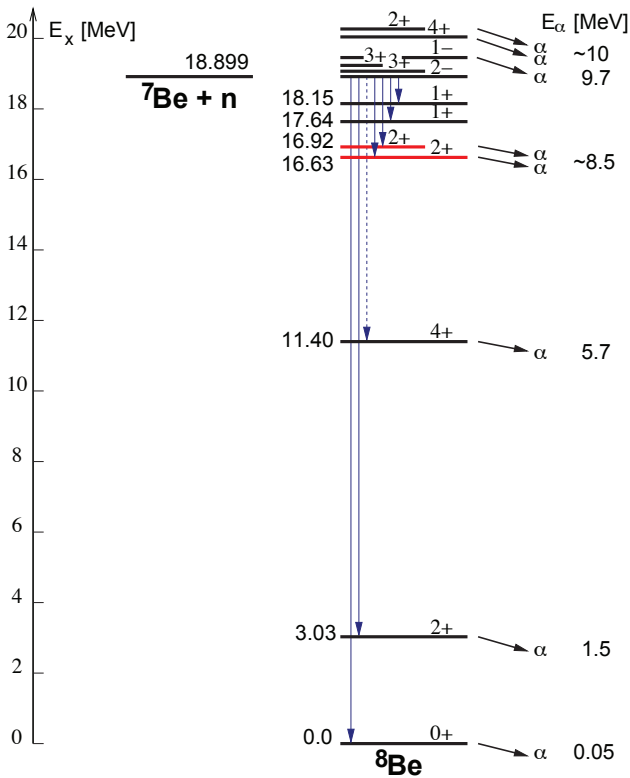}
\caption{Schematic level and decay scheme of $^8$Be showing
the states relevant for this measurement and related calculations \cite{levels}. The levels accessible in the present measurement are indicated in red. \label{fig:drc-levels}}
\end{figure}

The lack of experimental data for this reaction is essentially due to the intrinsic difficulty of the measurement, related to the low cross section and to the extremely high specific activity of $^{7}$Be (13 GBq/$\mu$g). The recent construction of a second, high-flux experimental area at n\_TOF (EAR2), CERN \cite{chiave}, offered the unique opportunity to perform a first time-of-flight measurement of the  $^{7}$Be($n, \alpha$)$^{4}$He cross section over a wider energy range. The neutron beam in the new measuring station, located 20 m above the spallation target  \cite{christina}, is characterized by an extremely high luminosity, wide energy range, good energy resolution and low repetition rate. All these features make EAR2 ideal for measurements on isotopes only available in very small amounts, with short half-lives, or both, as indeed is the case for $^{7}$Be. Two more conditions had to be met in order to perform the measurement: the availability of a $\mu$g-sized chemically pure $^{7}$Be sample and a highly efficient experimental setup. The measurement relied on the coincident detection of both $\alpha$-particles emitted in the $^{7}$Be($n, \gamma\alpha$)$^{4}$He reaction. As shown in Figure \ref{fig:drc-levels} and discussed later on, the reaction proceeds through various energy levels of $^{8}$Be populated by $\gamma$-ray transitions. However, the channels involving the ground and first exited state result in the emission of low-energy $\alpha$-particles ($\leq$1.5 MeV each), which could not be detected in the enormous $\gamma$-ray background of the $^7$Be sample. Therefore, only the partial cross sections involving the highest allowed excited levels between 16.6 and 18.1 MeV, which lead to the emission of two $\alpha$-particles with energy in excess of 8 MeV each, are accessible experimentally. The setup consisted of two Si-$^{7}$Be-Si sandwiches placed directly in the neutron beam. The Si detectors were 3$\times$3 cm$^{2}$ in area and 140 $\mu$m in thickness. A detailed description of the experimental setup can be found in Ref. \cite{cosentino,cemento}. 

The samples were prepared by the Paul Scherrer Institute (PSI), Switzerland. Thanks to the high selectivity of the experimental setup to the ($n, \alpha$) reaction, no isotopic separation was required, but to minimize the background from neutron-induced reactions on $^{7}$Li, the sample material was chemically purified a few days before the measurement. Two samples 3 cm in diameter were prepared with $\sim$1.4 $\mu$g $^{7}$Be each (corresponding to an activity of $\sim$18 GBq), one by electrodeposition on a 5-$\mu$m-thick Al foil and one by droplet deposition on a 0.6-$\mu$m-thick low-density polyethylene foil. The amount of $^{7}$Be on the samples was determined by their $\gamma$-ray activity.  

The two Si-$^{7}$Be-Si arrays were mounted in an aluminium chamber with thin entrance and exit windows for the neutron beam. On the side, the chamber was shielded by 1 cm of Pb for radioprotection reasons. A software reconstruction routine \cite{petar} was applied to the digitized signals to extract amplitude and time information. The energy calibration of the detectors was performed by means of the $^{6}$Li($n, t$)$^{4}$He reaction, measured with the final detector configuration a few days before the insertion of the $^{7}$Be samples. The same measurement provided an indication of the upper neutron energy limit of the measurement, which was determined by the recovery time of the Si detectors after the $\gamma$-flash from the spallation target. Therefore, the present results are limited to neutron energies below 10 keV. A time window of $\pm$100 ns was allowed for $\alpha$-$\alpha$ coincidences. The efficiency of the setup for these coincidences was estimated to be 40 and 37\% for the electrodeposited and droplet samples, respectively (due to a slightly different spatial distribution of $^{7}$Be in the two samples) \cite{cosentino}. Since the samples are slightly smaller than the neutron beam spatial profile, data were corrected for the beam interception factor, of 70$\pm$3\%.

The main background component in the measurement is related to the $\gamma$-rays from the $^{7}$Be decay. Although the individual signals correspond only to a few tens of keV, the large pile-up probability due to the very high activity could produce signals that mimic a high-energy deposition event. A similar argument applies to the pile-up of protons from the competing $^{7}$Be($n, p$)$^{7}$Li reaction, characterized by a very large cross section at thermal neutron energy ($\sim$39 kb). This background component was significantly reduced applying a 2 MeV threshold on the signal amplitude. The random coincidences produced by residual pile-up events were estimated via the coincident signals of uncorrelated Si detectors (i.e. belonging to different arrays).

Another potential source of background is the production of $^{8}$Li via neutron capture on $^{7}$Li, which undergoes $\beta$-decay into $^{8}$Be with a half-life of 800 ms. Since $^{7}$Li  builds up in the sample by $^{7}$Be decay, the contribution of this background component increases as a function of time (by the end of the measurement, which lasted 50 days, almost half of the $^{7}$Be had decayed into $^{7}$Li). As only the 3.03 MeV state of $^{8}$Be is populated by $^{8}$Li decay, the $\alpha$-particles are emitted with an energy of 1.5 MeV, well below the threshold used in the present experiment. Moreover, this type of background is suppressed by more than two orders of magnitute thanks to the very low duty cycle of the n\_TOF neutron beam.

Other sources of background are related to reactions induced by neutrons with energies above a few MeV, in particular by $^{9}$Be($n, 2n$), as well as the $^{7}$Li($p, \gamma$) and $^{7}$Be($p, \gamma$), with protons produced by neutron scattering in the polyethylene backing. All these reactions lead to $^{8}$Be, and thus to $\alpha-\alpha$ coincidences in the final state. However, they are easily discriminated via neutron time-of-flight. 

\begin{figure}[h]
\includegraphics[angle=0,width=1.\linewidth,keepaspectratio]{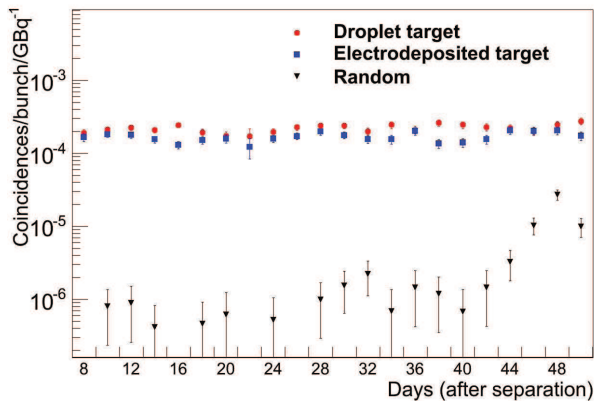}
\caption{(Color online) Detected number of coincidences as a function of time during the measurement. The results from the two Si-$^7$Be-Si arrays are indicated by red circles and blue squares. The background due to random coincidences averaged over all possible combinations of uncorrelated detectors is indicated by the black triangles. \label{fig:two}}
\end{figure}

\begin{figure}[t]
\includegraphics[angle=0,width=1.05\linewidth,keepaspectratio]{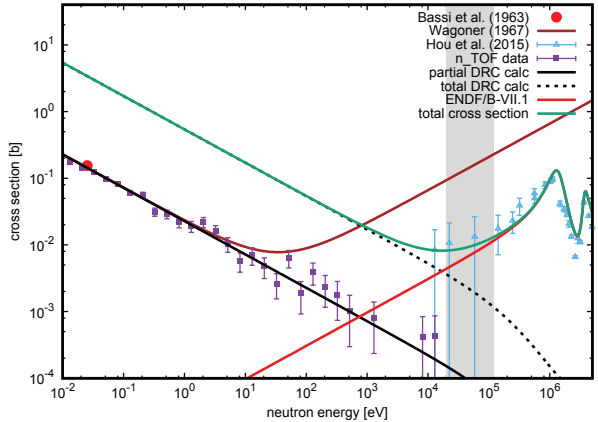}
\caption{\label{fig:sig-drc} (Color online) The partial $^{7}$Be(n,$\gamma\alpha$) cross section 
measured at n\_TOF between 10 meV and 10 keV (blue squares), 
compared with the result of Bassi et al. \cite{bassi} at thermal energy (red circle). 
The solid black curve represents the DRC calculations of the partial cross section 
for the excited states in $^{8}$Be 
between 16.6 and 18.2 MeV accessible in the present experiment, 
while the dashed curve shows the total cross section obtained by considering 
the branching ratios between different transitions. 
The cross section suggested in Ref. \cite{wagoner} 
and used in BBN calculations is shown by the brown curve. A combination of the present results with ENDF/B-VII.1 evaluation based on indirect measurements yields the total cross section up to a few MeV (green curve). The hatched area shows the neutron energy region of interest for Big Bang Nucleosyntesis.}
\end{figure}

Figure \ref{fig:two} shows the number of recorded counts (integrated over two days of beam time and normalized for the time-dependent $^{7}$Be content in both samples) over the whole duration of the measurement (red circles and blue squares). The error bars represent the statistical uncertainties only, while the systematic uncertainty on each dataset, essentially related to the determination of the sample mass, is of the order of 10\%. Within statistical uncertainties, the count-rates of both Si-$^{7}$Be-Si arrays are constant in time, excluding any background related to the $^{7}$Li build-up, which is expected to produce a positive slope in the number of counts. The background generated by random coincidences is small and well under control (black triangles). The consistent results obtained with the two arrays, within the respective uncertainties, demonstrate that the sample backing as well as the deposition technique had no visible effect on the overall detection efficiency and the related background. Figure \ref{fig:sig-drc} shows the measured cross section, obtained as a weighted average of the results from both arrays, yielding an overall systematic uncertainty slightly below 10\%.

The correct interpretation of the results requires some preliminary considerations. 
Low energy (s-wave) neutron interactions with $^{7}$Be in its ground state are strongly 
affected by the $J^{\pi}=2^{-}$ state in $^{8}$Be at 18.91 MeV excitation energy, 
just above the neutron separation energy of 18.899 MeV.  
The ($n, p$) channel is strongly enhanced by this state, 
resulting in a cross section for thermal neutrons of over 39 kb. 
The ($n, \alpha$) channel, on the contrary, is strongly suppressed due to parity conservation, 
which forbids the direct breakup of this state into two $\alpha$ particles \cite{suppl}. 
Little is known experimentally of the ($n, \gamma$) channels. 
In their study of parity violation in the strong interaction, Bassi {\it et al.} \cite{bassi} 
succeeded in measuring the ($n, \gamma\alpha$) cross section at thermal neutron energies, 
whereas they could only obtain an upper limit for the ($n, \alpha$) channel. 
Our result confirms that the direct ($n,\alpha$) breakup channel 
is strongly suppressed and that the ($n, \gamma\alpha$) reaction 
remains the dominant process of $\alpha$ emission 
following neutron capture on $^{7}$Be, for all bound states of $^{8}$Be.

The present results shed also light on the role of the positive parity states 
above the neutron separation threshold (see Fig. \ref{fig:drc-levels}),
which can be formed by incident p-wave or higher, odd-l wave, neutrons. 
If any of those states (a $3^{+}$ doublet, the second $4^{+}$ 
and an additional $2^{+}$ state for excitation energies up to 20.1 MeV) 
strongly contributed to the $\alpha$-emitting cross section, 
the tails of the corresponding resonances should have caused an observable 
deviation from the 1/v behaviour from thermal up to the keV energy region.


\begin{table}
\caption{$^8$Be bound states for
$| ^{7}$Be$(3/2^{-})\otimes \nu p_{3/2} > $ single particle configurations and Woods-Saxon potential parameters used in the calculations of the $(n,\gamma\alpha)$ DRC cross section.
\label{tab:dc} }
\begin{ruledtabular}
\begin{tabular}{rcccc}
\hline
E$_{x}$ & $J^\pi$     &  C$^{2}$S     & Thermal cross & cross section \\
MeV     &             &               &  section (mb)           & fraction (\%)   \\
\hline
0.000  & 0$^+$         & 1.51         & 1278 & 37.6       \\
3.030  & 2$^+$         & 0.57         & 1977 & 58.2       \\
16.626 & 2$^+$         & 0.30         & 63   &  1.9       \\
16.922 & 2$^+$         & 0.47         & 80   &  2.3       \\
\end{tabular}
\begin{tabular}{lr}
 Radius parameter  & r$_0$ =1.236 fm
\\
 Diffuseness
& $d$ = 0.62 fm  \\
 Spin-orbit strength & $V_{so}$ = 7.0 MeV \\  Well depth & $V_0$ = 60.3 MeV  \\ \end{tabular} \end{ruledtabular} \end{table}

As mentioned above, due to the constraints on the accessible $\alpha$ energy range, 
only electromagnetic dipole (E1) transitions feeding the group of $^{8}$Be states 
between 16.626 and 18.150 MeV excitation energy are relevant for the present study. However, because of parity conservation, which forbids the decay of the two J$^{\pi}$=1$^{+}$ states 
at 17.64 and 18.15 MeV, only the two J$^{\pi}$=2$^{+}$ states at 16.626 and 16.922 MeV are contributing to the present data 
and to the thermal cross sections measured by Bassi \textit{et al.} \cite{bassi}.
The remaining components of the reaction cross section can be derived 
from theoretical calculations via the direct radiative capture (DRC) mechanism, 
i.e. direct $\gamma$ decays from the capturing states 
to the relevant states in $^8$Be at 11.350 MeV and below (Fig. \ref{fig:drc-levels}). 
The corresponding DRC cross sections were calculated 
according to the prescription in Ref. \cite{Alberto} 
by determining the overlap integral of the bound-state 
with the continuum wave functions
obtained with the set of  Wood-Saxon parameters listed in Table \ref{tab:dc}. 
The spectroscopic factors (C$^{2}$S) have been derived by OXBASH \cite{oxbash} 
shell model calculations with the Kumar hamiltonian \cite{kumar} 
that are reproducing the level scheme of $^8$Be 
up to the 3$^+$ levels at 19.07 and 19.235 MeV fairly well.
It has to be noted here that the $1^{-}$ state at 19.4 MeV can be populated by s-wave neutrons, 
and could affect the present data by its 1/v resonance tail, 
complementing that of the $2^{-}$ state just above the threshold. 
A direct breackup into two $\alpha$ particles from the 1$^{-}$ 
state can be excluded on the basis of Ref. \cite{bassi}. 
However, a $(n,\gamma\alpha)$ process which includess the initial $1^{-}$ state  
needs to be considered and has been, in fact, included 
in our DRC modeling of the reaction process. 

While the calculated cross sections depend strongly on the choice of the strength of the interaction potential, 
the fractional contributions of the DRC components can be reliably determined, 
as they depend only weakly on these parameters. 
Therefore, the total $^{7}$Be($n, \gamma\alpha$) cross section 
can be obtained by combining the experimental 
data with the calculated fractional contributions. 
The result is shown in Fig. \ref{fig:sig-drc} by the dashed black curve. 
It exceeds the sum of the minor components (normalized to the experimental data) 
by more than a factor of 20, in striking disagreement 
with the previous estimate of Ref. \cite{wagoner}, 
in magnitude as well as in cross section shape.

To complete the picture of the $^{7}$Be($n, \alpha$) reaction at higher energy, cross section data derived from indirect reactions, namely $^{4}$He($\alpha, n$)$^{7}$Be and $^{4}$He($\alpha, p$)$^{7}$Li, recently published in \cite{hou} are shown in Fig. \ref{fig:sig-drc} (light blue triangles), together with the cross section from the evaluated nuclear data library ENDF/B-VII.1 (red curve). The ENDF evaluation has been based on a R-matrix analysis of several indirect reactions, which included the direct breakup of the positive-parity states with even angular momentum above the neutron separation threshold, such as the 2$^{+}$ states at 20.1 and 22.4 MeV. Since these levels can only be populated by p-wave neutrons, the corresponding resonant cross section increases with neutron velocity, as originally assumed in Ref. \cite{wagoner}, although the absolute value turns out to be incompatible with the present data.

The present results allow one to draw some conclusions 
on the role of the $^{7}$Be($n, \alpha$)$^{4}$He reaction for the Cosmological Lithium Problem. 
Regarding the trend of the cross section as a function of neutron energy (Fig. \ref{fig:sig-drc}) 
the experimental data are clearly incompatible with the theoretical 
estimate used in BBN calculations \cite{wagoner}, 
which was assuming that the cross section is determined 
at low energies by an s-wave entrance channel producing a 1/v behavior, 
and that it is dominated above a few keV by a p-wave channel, 
causing an increase with neutron velocity. Both components were constrained 
by the measured thermal values of Ref. \cite{bassi}. 
The present data indicate that both assumptions have to be reconsidered. In fact, a combination of the present results for the total ($n, \gamma\alpha$) cross section with the ENDF/B-VII.1 evaluation yields the overall energy dependent cross section shown by the green curve in Fig. \ref{fig:sig-drc}, sensibly different from that of Ref. \cite{wagoner}. The numerical tabulated cross sections shown in the figure, which can be used to make a new estimate of the reaction rate, can be found in the Supplemental Material \cite{suppl}.

In summary, the energy-dependent $^{7}$Be($n, \gamma\alpha$) cross section 
has been measured for the first time over a wide neutron energy range 
in the high-flux experimental area (EAR2) at n\_TOF. 
The main difficulties that have prevented the measurement of this cross section 
so far have been overcome thanks to the extremely high luminosity of the neutron beam, 
the extraction, chemical separation, deposition, and characterization 
of a sufficient amount of $^{7}$Be, and, finally, 
by a suitable experimental setup based on Silicon detectors. 
The total reaction cross section up to 10 keV obtained by the present results 
in combination with theoretical estimates indicate that 
below E$_{n}$$\approx$100 eV the cross section is over a factor of 20 higher 
than the one used in BBN calculations so far \cite{wagoner}. 
The 1/v behavior of the cross section observed up to the maximum energy of this measurement 
is excluding a significant p-wave contribution in contrast to the assumption of Ref. \cite{wagoner}. 
Although a final word on this reaction and its influence on the Cosmological Lithium Problem 
requires additional measurements extending to the MeV region, these first results 
provide evidence that the reaction rate currently used in BBN calculations 
requires substantial revision.
As a final remark, we would like to point out that the present results 
demonstrate the suitability of the recently constructed  n\_TOF second experimental area 
for performing challenging neutron measurements on unstable isotopes of short half life, 
of reactions characterized by a low cross section, with samples of extremely small mass. 
Indeed, for all three instances at once,
this has been the case of the present $^{7}$Be($n, \alpha$) measurement.

\begin{acknowledgments}
The authors wish to thank the PSI crew A. Hagel, D. Viol, R. Erne, K. Geissmann, O. Morath, F. Heinrich and B. Blau for the performance of the $^{7}$Be collection at the SINQ cooling system. This research was partially funded by the European Atomic Energy Communitys (Euratom) Seventh Framework Programme FP7/2007-2011 under the Project CHANDA (Grant No. 605203). We
acknowledge the support by the Narodowe Centrum Nauki (NCN), under the grant UMO-2012/04/M/ST2/00700.
\end{acknowledgments}

\bibliography{7Bena}

\end{document}